\documentclass[a4paper,11pt]{article}
\usepackage{jheppub} 
\usepackage{lineno}
\usepackage{float}
\usepackage{mathtools}
\usepackage{amsmath}
\usepackage{amsthm}
\usepackage{amssymb}
\usepackage{esint}
\usepackage[numbers]{natbib}

\usepackage[T1]{fontenc} 
\usepackage{amssymb}
\usepackage{stmaryrd}
\usepackage{tensor, physics, cancel, amsmath,amsfonts, mathtools}
\usepackage{graphicx}
\usepackage{dsfont}
\usepackage{xcolor}
\usepackage{comment}

\makeatletter
\def\@fpheader{\relax}
\makeatother

\newcommand{\Rmnum}[1]{\uppercase\expandafter{\romannumeral #1\relax}}

\newcommand{\be}{\begin{equation}}
\newcommand{\ee}{\end{equation}}

\newcommand{\opo}{\mathcal{O}}

\title{Estimating Quantum Gravity Corrections to Correlators near Black Holes}

\author{Ben Freivogel$^1$ and Tianyi Li$^{1,2}$}
\affiliation{$^1$ ITFA and GRAPPA\\
Universiteit van Amsterdam, Netherlands}
\affiliation{$^2$ Department of Physics,\\ University of Wisconsin-Madison, Madison, WI 53706, U.S.A.}

\abstract{We analyze the size of quantum gravity effects near black hole horizons. By considering black holes in asymptotically AdS spacetime, we can make use of the `quantum deviation' to estimate the size of quantum gravity corrections to the semiclassical analysis. We find that, in a typical pure state, corrections to correlation functions are typically of order exp(-S/2). Both the magnitude and time dependence of the correlator differ from previous results related to the spectral form factor, which estimated the correlator in a thermal state. Our results severely constrain proposals, such as non-violent unitarization and some versions of fuzzballs, that predict significant corrections to the semiclassical computation of correlation functions near black holes. We point out one possible loophole: our results rely on the standard result that bulk reconstruction is state independent for small perturbations outside the black hole.}

\begin{document}
\maketitle

\section{Introduction}\label{sec:intro}
The black hole information problem demonstrated that quantum gravity effects can become important at distance scales much longer than the Planck scale in the vicinity of black holes, raising the possibility that they may be observable. Recently, significant progress has been made in computing the quantum gravity corrections that prevent information loss (see for example \cite{Almheiri:2020cfm} for a review and references.)
Although quantum gravity effects are large in the computation of the entropy of the Hawking radiation, the entropy is not an observable quantity. 

It is an open question whether quantum gravity effects are significant for observable quantities, such as correlation functions of gravitons and other light fields. In some models, such as non-violent unitarization \cite{Giddings:2012gc, Giddings:2016plq, Giddings:2017jts, Giddings:2019vvj} and versions of the fuzzball proposal \cite{Mathur:2008nj, Bena:2013dka}, quantum corrections to correlators are significant near the horizon. Concretely, in the simplest version of non-violent unitarization, the two-point function of the gravitational field differs at order one from the semiclassical computation at distance and time scales of order the Schwarzschild radius. In non-violent unitarization, the two-point function of the metric perturbation in a typical pure state depends strongly on the state; this is essential for information to get out of the black hole in this model.

On the other hand, it is widely believed that corrections to correlation functions are exponentially small in the black hole entropy. Corrections of this order are sufficient to restore unitarity. In this paper we will compute the size of these corrections for black holes in asymptotically AdS spacetime, strongly constraining fuzzballs and nonviolent unitarization.

The central question we seek to address is whether the microstates of a black hole can be probed or inferred by examining quantum operators outside the event horizon. Specifically, we ask: to what extent is an operator $O$ outside the event horizon sensitive to these quantum microstates? 

To calculate this, we use the notion of `quantum deviation' of an operator $O$, defined by \cite{Freivogel:2021ivu}
\begin{equation}
\Delta_O^2=\int d \psi f(\psi)\left(|\langle\psi|O| \psi\rangle|^2\right)-\left(\int d \psi f(\psi)|\langle\psi|O| \psi\rangle|\right)^2,
\end{equation}
where $f(\psi)$ denotes a probability distribution over an ensemble of states. This $\Delta_O$ is designed to capture the average fluctuations, in an ensemble of states, in the {\it expectation value} of the operator $O$. It differs from the more standard quantity $\bra{\psi}O^2\ket{\psi}$; our quantity captures to what extent $\bra{\psi}O\ket{\psi}$ varies over some ensemble of states $\ket{\psi}$. As a result, this quantity is perfectly suited to asking to what extent operators outside the black hole are sensitive to the black hole microstate. 

\subsection{Summary of Results}
\paragraph{Magnitude.}
We find that quantum corrections are of order
\begin{equation}
\Delta_O^2 \sim e^{- S}
\end{equation}
for low-point correlation functions, where $S$ is the Bekenstein-Hawking entropy of the black hole. This corresponds to a typical difference in correlators of
\begin{equation}
\bra{\psi} O \ket{\psi} - \overline{\bra{\psi} O \ket{\psi}} \sim e^{-S/2}
\end{equation}
where the barred quantity is the ensemble average.
A particular case of interest to the black hole information problem is where $O$ is a long-time correlator, $O = A^\dagger(t) A(0)$ \cite{Maldacena:2001kr}. In this case our result gives
\begin{equation}
\bra{\psi} A^\dagger(t) A(0) \ket{\psi} \sim e^{-S/2}
\end{equation}
This should be contrasted with computations related to the spectral form factor, such as \cite{Cotler:2016fpe}, which give
\begin{equation}
    \expval{A^\dagger(t) A(0)}_\beta \sim e^{-S}
\end{equation}
This is not in conflict: the correlator in a typical pure state is much larger than the thermal correlator, albeit still exponentially small.
\paragraph{Time Dependence.}
We define a generalization of the quantum deviation in section \ref{sec-2}, the `quantum covariance.' We make use of the quantum covariance to diagnose the typical time dependence of the long-time correlator. Schematically, we find
\be
\overline{\expval{A(t) A(0)}\expval{A(t') A(0)}^*} \sim e^{-S}\expval{A_L(t) A_R(t')}_\beta 
\ee
where the bar again denotes the ensemble average.

This result shows that the correlation time is given by the thermal correlator, and is thus of order $\beta$ in simple examples. In other words, the value (magnitude and phase) of the long-time correlator changes on a typical timescale $\beta$. This should be contrasted with the wild oscillations of the thermal long-time correlator \cite{Cotler:2016fpe}, which occur on a time scales much shorter than $\beta$.

\paragraph{Regime of Validity.}
We demonstrate that the key formula, which relates the quantum deviation to a two-sided correlator in the thermofield double state, is reliable for any operator as long as the entropy is large.

Our black hole computation is special in various ways- we focus on particular dimensions and particular modes of the gravitational fluctuations. We expect that our results will go through more generally, but have not shown this rigorously.

\paragraph{Reconstruction of the black hole metric.}
In order to relate our quantum deviation formula, which applies to CFT operators, to bulk physics, we need to make use of the bulk-boundary dictionary. We show, applying previous results in the literature, the explicit formula relating the bulk gravitational fluctuations to the fluctuations in the boundary stress tensor.

\paragraph{Magnitude of the quantum deviation.}
We apply our `bulk reconstruction' formula to compute the quantum deviation of the scalar channel perturbation for a 5-dimensional AdS planar black hole. We estimate the order of magnitude of this quantity and find it must be exponentially small in the entropy. Our result is
\begin{equation}
\Delta_{g_{\mu \nu}}^2 \sim  e^{-S} e^{D \tilde{q} }  \quad \quad \tilde{q} \gg 1
\end{equation}
where $\tilde q$ is the momentum of the mode in units of the temperature, $D \approx 2.6$ is a constant, and the above formula is valid when the momentum is large compared to the frequency.

Formally, the above formula would allow for order one quantum deviation for sufficiently large momentum, but imposing a Planck-scale cutoff on the momentum imposes that $\tilde q \ll S$, so that the final answer is simply
\be
\Delta_{g_{\mu \nu}}^2 \sim  e^{-S} ~.
\ee
Our full result is presented in section \ref{sec-bh}.

Higher point functions can have larger quantum deviation: the quantum deviation of an n-point correlator is given by
\be
\Delta_{g^n}^2 \sim e^{-S} e^{n D \tilde q}
\ee
This formula gives quantitative support for the conventional wisdom that corrections to low-point functions are exponentially small, while corrections to high-point functions become order one, as required to solve the black hole information problem. The above formula is reliable for $n$ sufficiently small. It is tempting to extrapolate to large $n$ but in this regime corrections to the bulk-boundary mapping become important, so further work is needed.

\subsection{Discussion}
We discuss some possible loopholes in our constraints. In particular, our results rely on the state-independent reconstruction of operators outside the horizon in the context of AdS/CFT. This type of reconstruction is standard for physics outside the horizon. One way to escape our conclusions is to claim that the same quantum corrections posited in non-violent unitarization result in state-dependent modifications of the bulk-boundary dictionary. 

Our results are closely related to `Critique of the Fuzzball Program' \cite{Raju:2018xue} of Raju and Shrivastava, which also makes use of the quantum deviation, and uses it to argue that quantum gravity effects must be smaller than of order $\exp(-S/2)$. We make two main additions to that work. First, we give an estimate rather than simply a bound, showing that in fact quantum gravity effects must be of order $\exp(-S/2)$ in pure states. Second, their critique relies on the concept of an ensemble of black hole states. It is not clear how well-defined this is in the bulk gravitational theory, particularly in asymptotically flat spacetime. We make this ensemble well-defined by making use of AdS/CFT. We then clarify, through a detailed examination of the bulk-boundary mapping, how to place bounds on bulk observables. Embedding the discussion in AdS/CFT not only makes the ensemble more well-defined; it also makes clear the possible loophole coming from the bulk-boundary mapping described in the previous paragraph.

{\bf Note:} This work originated as the masters thesis of TL \cite{thesis}. The full version of the thesis contains much more background material but not all of the new results of this paper.

\section{Quantum Deviation and Quantum Covariance}
\label{sec-2}

Quantum deviation is a core concept in this paper, used to measure the fluctuation in the {\it expectation value} of an operator in an ensemble of states. Note that this is very different from the usual variance of an operator: the quantum deviation quantifies the spread in expectation values $\expval{\opo}_i$ in some ensemble of states $\ket{i}$, rather than the variance $\expval{\opo^2}$ in some given state. Our goal is to probe the sensitivity of metric fluctuations near the horizon of a black hole to the microstate of the black hole. While we cannot compute the expectation value in one given microstate, the quantum deviation allows us to compute the typical deviation in an ensemble of microstates.  

In the following sections, we will introduce the concept of quantum deviation, as well as a formula for calculating the quantum deviation of an operator in a canonical ensemble, which is manifested as a connected two-point correlator of the operator in a thermofield double state. We will also justify the validity of the approximations used in the derivation of this formula. We also introduce a generalization of the quantum deviation which we name the {\it quantum covariance}. This quantity will turn out to be very useful in diagnosing the behavior of long-time correlators.

\subsection{Review of quantum deviation}\label{roqd}
The quantum deviation is a measure of the fluctuations in an expectation value of an operator in an ensemble of pure states.  This concept is introduced in \cite{Freivogel:2021ivu}. The quantum deviation and the usual variance of a quantum operator are quite different. The regular variance, written as $\langle O^2 \rangle - \langle O \rangle^2$, tells us how much measurement results spread out in a specific quantum state. On the other hand, the quantum deviation shows us how the expected value wiggles across a bunch of states. While we can find the regular variance using the system's density matrix, the quantum deviation only makes sense when we've picked out a way to distribute probabilities among different quantum states. So, $\Delta_O$ works well for figuring out if normal black holes in pure quantum states show any horizon structures that vanish after we do a bunch of state-averaging. 

Consider an ensemble of states $\rho$ with associated probabilities $P(\rho)$. One can perform experiments on multiple copies at once. Given such an ensemble, the quantum deviation $\Delta_O$ of an operator $O$ is defined as \cite{Freivogel:2021ivu}
\begin{equation}
	\Delta^2_O\equiv \Bigl\llbracket \bigl\vert\Tr(O\rho)\bigr\vert^{2}\Bigr\rrbracket- \Bigl\vert\big\llbracket\Tr(O\rho)\bigr\rrbracket\Bigr\vert^{2}.
\end{equation}
Here the double bracket represents averaging over an ensemble of states, $\llbracket \cdot \rrbracket=\sum_{ \rho } P( \rho )$. Note that $\rho$ denotes a state within the ensemble instead of the ensemble itself. If we consider an ensemble of pure states by letting $\rho=\ket{\psi}\bra{\psi}$, the quantum deviation is given by
\begin{equation}
    \Delta^2_O\equiv \Bigl\llbracket \bigl\vert\bra{\psi}O\ket{\psi}\bigr\vert^{2}\Bigr\rrbracket- \Bigl\vert\bigl\llbracket\bra{\psi}O\ket{\psi}\bigr\rrbracket\Bigr\vert^{2}.
\end{equation}
To make the concept of quantum deviation more clear, we follow \cite{Freivogel:2021ivu} to study a simple example to illustrate how the information is carried by the quantum deviation. We describe this example in detail for clarity; the impatient reader can jump to the next subsection.

Let's focus on two distinct ensembles for a spin-${1 \over 2}$ particle. In Ensemble $A$, every pure state holds equal probability:
\begin{equation}
    A\equiv\Bigl\{\Bigl(\ket{\hat{n}}, p_{\hat{n}}=\frac{1}{4\pi}\Bigr)\Bigr\}_{\hat{n}\in S_2}\,,
\end{equation}
where $\hat{n}$ denotes a point on the Bloch sphere and
\begin{equation}
\ket{\hat{n}} = \cos{\frac{\theta}{2}}\ket{\uparrow}+e^{i\phi}\sin{\frac{\theta}{2}}\ket{\downarrow},.
\end{equation}
On the other hand, Ensemble $B$ has the particle in the $\ket{\uparrow}$ state with probability 1/2 and in the $\ket{\downarrow}$ state with probability 1/2: 
\begin{equation}
    B\equiv\Bigl\{\Bigl(\ket{\uparrow}, p_{\uparrow}=\frac{1}{2}\Bigr),\Bigl(\ket{\downarrow}, p_{\downarrow}=\frac{1}{2}\Bigr)\Bigr\}\,.
\end{equation}
Let's take $O=\sigma_z,$ where $\sigma_z\ket{\uparrow}=1$ and $\sigma_z\ket{\downarrow}=-1.$ It's easy to compute the average expectations of this operator in both ensembles. For Ensemble $A,$ we find:
\begin{equation}
A:\quad\bigl\llbracket \langle\sigma_z\rangle\bigr\rrbracket=\frac{1}{4\pi}\int d\Omega_2 \bra{\hat{n}}\sigma_z\ket{\hat{n}}=0,,
\end{equation}
with $d\Omega_2$ being the volume element on the Bloch sphere. For Ensemble $B,$ we have:
\begin{equation}
B:\quad \bigl\llbracket \langle\sigma_z\rangle\bigr\rrbracket=\frac{1}{2}\bigl(\bra{\uparrow}\sigma_z\ket{\uparrow}+\bra{\downarrow}\sigma_z\ket{\downarrow}\bigr)=0,.
\end{equation}
So, the average expectation value of $\sigma_z$ fails to tell the difference between the two ensembles. Yet, the quantum deviation can make it. For Ensemble $A,$ we get:
\begin{equation}
A:\quad\Delta_{\sigma_z}^2 = \frac{1}{4\pi}\int d\Omega_2 \bra{\hat{n}}\sigma_z\ket{\hat{n}}^2 = \frac{1}{3},.
\end{equation}
Meanwhile, for Ensemble $B,$ we find:
\begin{equation}
B:\quad \Delta_{\sigma_z}^2=\frac{1}{2}\bigl(\bra{\uparrow}\sigma_z\ket{\uparrow}^2+\bra{\downarrow}\sigma_z\ket{\downarrow}^2\bigr)=1,.
\end{equation}

We can even reformulate the quantum deviation in terms of an averaged state, at the expense of dealing with a duplicated theory. It takes the form:
\begin{equation}
\Bigl\llbracket|\operatorname{Tr}(\mathcal{O} \rho)|^2 \rrbracket=\operatorname{Tr}\left(\rho_2 \mathcal{O}^{\dagger} \otimes \mathcal{O}\right)
\end{equation}
where a normalized density matrix in a duplicated Hilbert space is introduced, $\rho_2 \equiv \llbracket \rho \otimes \rho \rrbracket$. The quantum deviation is then expressed as:
\begin{equation}
\Delta_O^2=\operatorname{Tr}\left(\rho_2 \mathcal{O}^{\dagger} \otimes \mathcal{O}\right)-\left|\operatorname{Tr}\left(\rho_1 \mathcal{O}\right)\right|^2.
\end{equation}
It's apparent that tracing out either side in $\rho_2$ yields $\rho_1.$ Thus, the subtracted term in the above expression merely removes the disconnected part of the correlator, making the final outcome a connected 2-point function in the duplicated theory:
\begin{equation}
\Delta_O^2=\operatorname{Tr}\left(\rho_2\left(\mathcal{O}^{\dagger}-\left\langle\mathcal{O}^{\dagger}\right\rangle\right) \otimes(\mathcal{O}-\langle\mathcal{O}\rangle)\right),
\end{equation}
where $\langle\mathcal{O}\rangle=\operatorname{Tr}\left(\rho_1 \mathcal{O}\right)$. Returning once more to the spin example, we can verify that the double state $\rho_2$ differs between the two ensembles. For Ensemble $A,$ it's:
\begin{equation}
A:\quad\rho_{2}=\frac{1}{4\pi}\int d\Omega_2 \ketbra{n,n}= \frac{1}{3}\bigl(\ketbra{\downarrow,\downarrow}+\ketbra{\uparrow,\uparrow}+\ketbra{\psi_+}\bigr),,
\end{equation}
with $\ket{\psi_+}\equiv\bigl(\ket{\uparrow,\downarrow}+\ket{\downarrow,\uparrow}\bigr)/\sqrt{2}.$
For Ensemble $B,$ we can obtain that:
\begin{equation}
B:\quad\rho_{2}=\frac{1}{2}\Bigl(\ketbra{\downarrow,\downarrow}+\ketbra{\uparrow,\uparrow}\Bigr),.
\end{equation}
So, distributions $A$ and $B$ result in the same $\rho_1$ but differing $\rho_2$'s. The quantum deviation can distinguish between $A$ and $B$ for any operator $O$ sensitive to this distinction, like $\sigma_z.$

\par In the simple example we've demonstrated, we show that the quantum deviation's computation involves a correlation function within a state in the duplicated theory.  We will see this scenario can be generalized to more cases. Now we would like to study the case when the averaged state of the ensemble is given by
\begin{equation}\label{state}
    \Bigl\llbracket \rho \Bigr\rrbracket = \sum_n \frac{\hat{p}_n}{Z_1} \ket{n}\bra{n},
\end{equation}
where $\hat{p}_n$ is an unnormalized probability distribution with normalization $Z_1=\sum_n \hat{p}_n$ and $\ket{n}$ are energy eigenstates. In the following we will denote the $k$-th normalization to be $Z_k=\sum_n (\hat{p}_n)^k$. It is shown in \cite{Freivogel:2021ivu} that the quantum deviation in the ensemble \ref{state} can be computed by correlation functions in thermofield-like states. We have
\begin{equation}\label{quantum_deviation_1}
    \Delta_O^2 = \frac{Z_2}{Z_1^2}\left( \bra{T_2}O_L O_R\ket{T_2}_c - \frac{Z_4}{Z_2^2} \bra{T_4}O_L O_R\ket{T_4} + ...\right),
\end{equation}
where the dots represent a small correction $\tilde{\varepsilon}_{nm}^{\scriptscriptstyle(1)}$. Here the thermofield-like states are defined as
\begin{equation}
    \ket{T_k} \equiv \frac{1}{\sqrt{Z_k}}\sum_n \hat{p}_n^{k/2}\ket{\Tilde{n}}\ket{n},
\end{equation}
where $\ket{\tilde{n}}=\Theta\ket{n}$ and $\Theta$ is an anti-unitary operator. We have also defined $O_L \equiv \Theta O \Theta^{\dagger} \otimes \mathbb{I}$ and $O_R \equiv \mathbb{I} \otimes O$. Note that $\left\langle \cdot \right\rangle_c$ denotes the connected two-point function
\begin{equation}
    \bra{T_2}O_L O_R\ket{T_2}_c \equiv \bra{T_2}(O_L-\left\langle O_L\right\rangle_2) (O_R-\left\langle O_R \right\rangle_2)\ket{T_2}.
\end{equation}
We consider the probability distribution such that the ensemble is dominated by a large number, $e^S$, of states. In this case, we have $Z_4/Z_2^2 \sim e^{-S}$, so we expect that the connected two-point function in $\ket{T_2}$ gives the dominant contribution. We have
\begin{equation}\label{final_formula}
    \Delta_O^2 = \frac{Z_2}{Z_1^2}\Bigl( \bra{T_2}O_L O_R\ket{T_2}_c + O(e^{-S}) \Bigr).
\end{equation}

\par Taking canonical ensemble as an example, we set $\hat{p}_n=e^{-\beta E_n}$. Then the state $\ket{T_2}$ is nothing but a thermofield double state with temperature $2\beta$. We can write the quantum deviation
\begin{equation}\label{canonical_form}
    \Delta_O^2 = \frac{Z(2\beta)}{Z(\beta)^2}\Bigl( \bra{\text{TFD}_{2\beta}}O_L O_R\ket{\text{TFD}_{2\beta}}_c + O(e^{-S}) \Bigr).
\end{equation}
Since the thermofield double state is dual to an eternal black hole at sufficiently large temperatures, we can calculate the quantum deviation holographically by considering a connected two-point correlator in the black hole background.

\subsection{Quantum Covariance}
One object of interest is the time dependence of the long time correlator
\begin{equation}
\bra{\psi} A^\dagger(t) A(0) \ket{\psi}
\end{equation}
We would like to know whether this quantity varies quickly, as is the case for the thermal long-time correlator. To address this, we would like to evaluate the correlation between $\bra{\psi} A^\dagger(t) A(0) \ket{\psi}$ and $\bra{\psi} A^\dagger(t') A(0) \ket{\psi}$. This suggests that we generalize the definition of the quantum deviation to the {\it quantum covariance}:
\be
\Bigl\llbracket  \opo_1 \opo_2 \Bigr\rrbracket  \equiv \int d\psi f(\psi) \bra{\psi} \opo_1 \ket{\psi} \bra{\psi} \opo_2 \ket{\psi}
\ee
This is the obvious generalization of the usual definition of the covariance for a probability distribution. However, note that it is a slightly exotic quantity from the point of view of quantum mechanics, just like the quantum deviation.  The reason for considering this quantity is that it can be computed reliably, and gives information about correlators in typical pure states.

We can also define the connected two-point function as usual,
\be
\Bigl\llbracket  \opo_1 \opo_2 \Bigr\rrbracket_c  \equiv \int d\psi f(\psi) \bra{\psi} \opo_1 \ket{\psi} \bra{\psi} \opo_2 \ket{\psi} - \Bigl\llbracket \opo_1 \Bigr\rrbracket \Bigl\llbracket \opo_2 \Bigr\rrbracket 
\ee
With these definitions, the quantum deviation is a special case,
\be 
\Delta_\opo^2 = \Bigl\llbracket 
\opo^\dagger 
\opo \Bigr\rrbracket_c
\ee
We can generalize our previous derivation to relate the quantum covariance to a correlator in a thermofield double type state:
\be\label{quantum_covariance}
\Bigl\llbracket  \opo_1 \opo_2 \Bigr\rrbracket_c = \frac{Z_2}{Z_1^2}\Bigl( \bra{T_2}O_1^L O_2^R\ket{T_2}_c + O(e^{-S}) \Bigr).
\end{equation}
where $\ket{T_2}$ is the thermofield double-like state defined above, and the prefactor is schematically 
\be
{Z_2 \over Z_1^2} \sim e^{-S}
\ee

\subsection{Regime of Validity}
In the derivation of the equation (\ref{canonical_form}), we used the approximation (\ref{final_formula}), which means the high order terms in (\ref{quantum_deviation_1})  are suppressed by $e^{-S}$. This approximation is also employed in the generalization to the formula of quantum covariance (\ref{quantum_covariance}). Now we would like to verify the validity of the approximation. To do this, it is convenient to split the quantum covariance into three pieces, $\llbracket  \opo_1 \opo_2 \rrbracket_c=\text{I}-\text{II}+\text{III}$, where
\begin{equation}
\begin{aligned}
    \text{I} &=\frac{Z_{2}}{Z_{1}^{2}}\bra{T_{2}}O_1^{L}O_2^{R}\ket{T_{2}}_{c}= \frac{1}{Z_1^2}\sum_{nm}\hat{p}_{n}\hat{p}_{m}O_{1nm}O_{2nm}- \frac{1}{Z_1^2Z_{2}}\sum_{nm}\hat{p}_{n}^{2}\hat{p}_{m}^{2}O_{1nn}O_{2mm}\,,\\
	\text{II}&=\frac{Z_{4}}{Z_{1}^{2}Z_{2}}\bra{T_{4}}O_1^{L}O_2^{R}\ket{T_{4}}=\frac{1}{Z_{1}^{2}Z_{2}}\sum_{nm}\hat{p}_{n}^{2}\hat{p}_{m}^{2}O_{1nm}O_{2nm}\,,\\
	\text{III}&=\sum_{nm}\tilde{\varepsilon}_{nm}^{\scriptscriptstyle(1)}\bigl(O_{1nm}O_{2nm}+O_{1nn}O_{2mm}\bigr)\,.
\end{aligned}
\end{equation}
We then further split these pieces into several terms, $\text{I}=\text{A} - \text{B}$, $\text{II}=\text{C}$, and $\text{III}=\text{D}+\text{E}$, where
\begin{equation}
\begin{aligned}
    \text{A}&=\frac{1}{Z_1^2}\sum_{nm}\hat{p}_{n}\hat{p}_{m}O_{1nm}O_{2nm}, \\
    \text{B}&=\frac{1}{Z_1^2Z_{2}}\sum_{nm}\hat{p}_{n}^{2}\hat{p}_{m}^{2}O_{1nn}O_{2mm},\\
    \text{C}&=\frac{1}{Z_{1}^{2}Z_{2}}\sum_{nm}\hat{p}_{n}^{2}\hat{p}_{m}^{2}\hat{p}_{n}\hat{p}_{m}O_{1nm}O_{2nm},\\
    \text{D}&=\sum_{nm}\tilde{\varepsilon}_{nm}^{\scriptscriptstyle(1)}\hat{p}_{n}\hat{p}_{m}O_{1nm}O_{2nm},\\
    \text{E}&=\sum_{nm}\tilde{\varepsilon}_{nm}^{\scriptscriptstyle(1)}O_{1nn}O_{2mm}.
\end{aligned}
\end{equation}
\par We do not know the precise expression of $\tilde{\varepsilon}_{nm}^{\scriptscriptstyle(1)}$, but we assume that it is exponentially suppressed in the entropy comparing to the leading non-gaussian piece, i.e.
\begin{equation}\label{condition_1}
    \tilde{\varepsilon}_{nm}^{\scriptscriptstyle(1)}\approx e^{-S}\frac{1}{Z_1^2Z_{2}}\hat{p}_{n}^{2}\hat{p}_{m}^{2}.
\end{equation}
Using (\ref{condition_1}), we can derive that
\begin{equation}\label{condition_2}
    \text{D}\approx e^{-S} \text{C}, \quad \text{E}\approx e^{-S} \text{B}.
\end{equation}
As we mentioned above, we focus on the case such that a large number, $e^S$, of states dominate the ensemble. Thus we can write
\begin{equation}
    \frac{\hat{p}_n^2 \hat{p}_m^2}{Z_2} \approx e^{-S} \hat{p}_n \hat{p}_m.
\end{equation}
This assumption leads us to find that
\begin{equation}\label{condition_2}
    \text{C} \approx e^{-S} \text{A}.
\end{equation}
The quantum covariance can then be rewritten as
\begin{equation}
\begin{split}
    \Bigl\llbracket  \opo_1 \opo_2 \Bigr\rrbracket_c &= \text{A}-\text{B} + \text{E} - \text{C} + \text{D}  \\
               &= (\text{A} - \text{B}) - e^{-S}(\text{A} - \text{B}) + e^{-2S}\text{A}\\
               &= \epsilon - e^{-S}\epsilon + e^{-2S}\text{A},
\end{split}
\end{equation}
where we have used (\ref{condition_1}) and (\ref{condition_2}) in the second equality and defined $\epsilon=\text{A}-\text{B}$ in the last line. Since the second term $e^{-S}\epsilon$ is exponentially suppressed in the entropy, it can always be neglected in the leading order approximation. We then have to discuss different cases of the relative sizes of $\epsilon$ and $\text{A}$. When $\epsilon \gg e^{-2S}\text{A}$, the first term dominates, and we have $\llbracket  \opo_1 \opo_2 \rrbracket_c \approx \epsilon = \text{A} - \text{B}$, which is exactly the two-point function (\ref{final_formula}). When $\epsilon$ is of the same order as $e^{-2S}\text{A}$, $\epsilon \approx e^{-2S}\text{A}$, we can write $\llbracket  \opo_1 \opo_2 \rrbracket_c \approx 2\epsilon = O(e^{-2S}\text{A})$. The last case is when $\epsilon \ll e^{-2S}\text{A}$, we have $\llbracket  \opo_1 \opo_2 \rrbracket_c \approx e^{-2S}\text{A}$. In summary, we can always write down
\begin{equation}
    \Bigl\llbracket  \opo_1 \opo_2 \Bigr\rrbracket_c = \text{A} - \text{B} + O(e^{-2S}\text{A}).
\end{equation}
\par It is obvious that both (\ref{final_formula}) and (\ref{quantum_covariance}) are correct if and only if $e^{-2S}\text{A}$ is negligible compared to $\text{A} - \text{B}$, which is expected to be of the case when we consider ensembles with very large entropy.

\subsection{Long-Time Correlators}

Now let's apply this formalism to determine the time dependence of the long-time correlator in a typical pure state. Let
\be
\opo_1 = \left( A^\dagger(t) A(0) \right)^\dagger \ \ \ \ \ \ \ \ \ \ \opo_2 = A^\dagger(t') A(0)
\ee
Then we have
\be
\Bigl\llbracket \bra{\psi} A^\dagger(t) A(0) \ket{\psi}^*\bra{\psi} A^\dagger(t') A(0) \ket {\psi} \Bigr\rrbracket  = {Z_2 \over Z_1^2}
\Bigl( \bra{T_2} A_L^\dagger(0) A_L(t) A^\dagger_R(t') A_R(0)\ket{T_2}_c + O(e^{-S}) \Bigr).
\end{equation}
Here we have used the Killing vector from the bulk to define the time, so that $t$ runs upwards in the right CFT and down in the left CFT. 

Using that the bulk theory is weakly coupled, the result is dominated by the two left-right Wick contractions (the same-side Wick contraction is not included because we are computing the connected part):
\begin{align}
&\Bigl\llbracket \bra{\psi} A^\dagger(t) A(0) \ket{\psi}^*\bra{\psi} A^\dagger(t') A(0) \ket {\psi} \Bigr\rrbracket  \approx \\
&{Z_2 \over Z_1^2}
\Bigl( \bra{T_2} A_L^\dagger(0) A^\dagger_R(t) \ket{T_2} \bra{T_2} A_L(t)  A_R(0)\ket{T_2} +  \bra{T_2} A_L^\dagger(0) A_R(0) \ket{T_2} \bra{T_2} A_L(t)  A^\dagger_R(t')\ket{T_2} + O(e^{-S}) \Bigr).
\end{align}
If we now take long times, $t, t' \gg \beta$, the first contraction becomes small, so we have
\be
\Bigl\llbracket \bra{\psi} A^\dagger(t) A(0) \ket{\psi}^*\bra{\psi} A^\dagger(t') A(0) \ket {\psi} \Bigr\rrbracket  \approx {Z_2 \over Z_1^2} \Bigl( \bra{T_2} A^\dagger_L(0) A_R(0) \ket{T_2} \bra{T_2} A_L(t)  A^\dagger_R(t')\ket{T_2} + O(e^{-S}) \Bigr).
\label{correl}
\end{equation}
This equation shows that the correlation between the long-time correlator at different times is proportional to the opposite-side correlator of the operator itself.

When $t=t'$, we can diagnose the typical size of $\bra{\psi} A^\dagger(t) A(0) \ket{\psi}$, as was done in \cite{Freivogel:2021ivu}:
\be
\Delta_{A^\dagger(t) A(0)}^2 \sim e^{-S} 
\ \ \ \ \ \to \ \ \ \ \ \ \ \ \ 
\bra{\psi} A^\dagger(t) A(0) 
\ket{\psi} \sim e^{-S/2}
\ee
This should be contrasted with the thermal result,
\be
\expval{A^\dagger(t) A(0)}_\beta \sim e^{-S}
\ee
Therefore, the long-time correlator is much larger in a typical pure state than in the thermal state, although both are exponentially suppressed in the entropy.

Having established the magnitude, we now want to diagnose the time dependence. This is determined by equation \eqref{correl} above,
\be 
\Bigl\llbracket \bra{\psi} A^\dagger(t) A(0) \ket{\psi}^*\bra{\psi} A^\dagger(t') A(0) \ket {\psi} \Bigr\rrbracket  \approx \Bigl( {Z_2 \over Z_1^2} \bra{T_2} A_L(0) A_R(0) \ket{T_2} \Bigr) \bra{T_2} A_L(t)  A^\dagger_R(t')\ket{T_2}
\ee 
The prefactor is independent of time, so the time dependence is captured by simply
\be
\boxed{
\Bigl\llbracket \bra{\psi} A^\dagger(t) A(0) \ket{\psi}^*\bra{\psi} A^\dagger(t') A(0) \ket {\psi} \Bigr\rrbracket  \sim e^{-S} \bra{T_2} A^\dagger_L(t)  A^\dagger_R(t')\ket{T_2}
}
\ee
Due to the symmetries, the correlator  $\bra{T_2} A_L(t) A^\dagger_R(t) \ket{T_2}$ is a function only of $t-t'$. (Recall that we are using the bulk Killing vector to define time, so that it goes `up' on the right boundary and `down' on the left boundary of the conformal diagram.) The precise behavior of this correlator depends on the details of the theory.
For example, consider a 2d CFT, with $A$
a scalar primary operator. Then
\be
\bra{T_2} A^\dagger_L(t)  A^\dagger_R(t')\ket{T_2} \sim {1 \over \cosh^{2 \Delta} (\beta (t - t'))}
\ee
So in simple cases, the characteristic correlation time of the long-time correlator is of order the inverse temperature $\beta$. This is in stark contrast to the thermal long-time correlator, which oscillates wildly on a shorter time scale.

\subsection{Check using ETH}
As a complementary way to understand and check the above results, we can estimate the behavior of the long-time correlator assuming that the operator $A$ satisfies the Eigenvector Thermalization Hypothesis (ETH). This is a special case; our analysis above is more general and does not assume ETH. We can even work directly from the definition of the quantum covariance, rather than making use of the results relating it to a correlator in the doubled theory. Parameterize the state in the energy eigenbasis
\be
\ket{\psi} = \sum_E \psi_E \ket{E}
\ee
Then
\begin{align}
&\Bigl\llbracket \bra{\psi} A^\dagger(t) A(0) \ket{\psi}^*\bra{\psi} A^\dagger(t') A(0) \ket {\psi} \Bigr\rrbracket  = \\
&\Bigl\llbracket \sum_{E_1 ... E_4} \psi(E_1) \psi^*(E_2) \bra{E_1} A^\dagger(t) A(0) \ket{E_2}^* \psi^*(E_3) \psi(E_4) \bra{E_3}A^\dagger(t') A(0) \ket{E_4} \Bigr\rrbracket
\end{align}
Inserting a complete set of states inside each correlator, this becomes 
\begin{align}
&\Bigl\llbracket \bra{\psi} A^\dagger(t) A(0) \ket{\psi}^*\bra{\psi} A^\dagger(t') A(0) \ket {\psi} \Bigr\rrbracket  = \\
&\Bigl\llbracket \sum_{E_1 ... E_6} \psi_{E_1} \psi^*_{E_2} \bra{E_1} A^\dagger(t) \ket{E_5}^* \bra{E_5} A(0) \ket{E_2}^* \psi^*_{E_3} \psi_{E_4} \bra{E_3}A^\dagger(t') \ket{E_6} \bra{E_6} A(0) \ket{E_4} \Bigr\rrbracket
\end{align}
Simplifying, this becomes
\be
\sum_{E_1 ... E_6} \Bigl\llbracket \psi_{1}\psi^*_{2}\psi^*_{3}\psi_{4} \Bigr\rrbracket e^{i (E_5 - E_1) t} e^{-i(E_6 - E_3) t'} (A^\dagger_{15})^* A_{52}^* A^\dagger_{36} A_{64}
\ee
where we have used the shorthand $
A_{15} \equiv \bra{E_1} A \ket{E_5}
$ and $\psi_1 \equiv \psi_{E_1}$. Using the definition of the Hermitean conjugate gives
\be
\sum_{E_1 ... E_6} \Bigl\llbracket \psi_{1}\psi^*_{2}\psi^*_{3}\psi_{4} \Bigr\rrbracket e^{i (E_5 - E_1) t} e^{-i(E_6 - E_3) t'} A_{51} A_{52}^* A^*_{63} A_{64}
\ee
Now we have a number of simplifications: from ETH, from large $t, t'$, and from the average over states. For $t$ large and $t$ near $t'$, in order for the phases not to kill the result, we need 
\be
E_5 - E_1 = E_6 - E_3
\ee
ETH gives a type of Wick contraction, requiring either
\be
E_1 = E_3 \ \ \ \ \ E_5 = E_6 \ \ \ \ \ E_2 = E_4
\ee
or
\be
E_1 = E_2 \ \ \ \ \ \ E_3 = E_4
\ee
The second of these options gives 
\be
\sum_{E_1, E_3} \Bigl\llbracket \psi_1 \psi^*_1 \psi^*_3 \psi_3 \Bigr\rrbracket \sum_{E_5} A_{51} A^*_{51} e^{i (E_5 - E_1)t} \sum_{E_6} A_{63} A^*_{63} e^{-i (E_6 - E_3)t'} 
\ee
This is a disconnected piece, equal to
\be
\sum_{E_1, E_3} \Bigl\llbracket \psi_1 \psi^*_1 \psi^*_3 \psi_3 \Bigr\rrbracket \bra{E_1} A^\dagger(t) A(0) \ket{E_1} \bra{E_3} A^\dagger(t') A(0) \ket{E_3}^*
\ee
Recall that the statistics of the $\psi_i$ in our ensemble of states are Gaussian to leading order,
\be
 \Bigl\llbracket \psi_1 \psi^*_1 \psi^*_3 \psi_3 \Bigr\rrbracket = p(E_1) p(E_3)(1 + \mathcal{O}(e^{-S}))
 \ee
 where $p(E)$ is the probability distribution over energies specifying our ensemble of pure states; $p(E)$ can be taken to be canonical or microcanonical, or something more exotic.

 Making use of this result, 
 we find that this contraction is simply two copies of the thermal long-time correlator,
 \be
 \sum_{E} p(E) \bra{E} A^\dagger(t) A(0) \ket{E} \times 
 \sum_{E'} p(E') \bra{E'} A^\dagger(t') A(0) \ket{E'}^* \approx \expval{A^\dagger(t) A(0)}_\beta \expval{A^\dagger(t') A(0)}^*_\beta
\ee
Using previous results on long-time correlators, this term is of order $\exp(-2S)$. We will see that it can be neglected because it is much smaller than the other term, which we now compute.

The other contraction gives
\be
\sum_{E_1, E_2, E_5} \Bigl\llbracket{\psi_1 \psi^*_2 \psi^*_1 \psi_2} \Bigr\rrbracket  e^{i(E_5 - E_1)(t - t')} |A_{15} |^2 |A_{52}|^2
\ee
Note that at this stage we can already see that the correlator only depends on the time difference.

Now we use that the statistics of $\psi_i$ is Gaussian to leading order, and that
\be
\Bigl\llbracket \psi_1^* \psi_2 \Bigr\rrbracket = \delta_{12} p(E_1) \ \ \ \ \ \ \ \ \Bigl\llbracket \psi_1 \psi_2 \Bigr\rrbracket = 0
\ee
Plugging this in gives
\be
\sum_{E_1, E_2, E_5} p(E_1) p(E_2) e^{i(E_5 - E_1)(t-t')}|A_{15} |^2 |A_{52}|^2
\ee
To make progress, we want to estimate the sum over $E_2$,
\be
\sum_{E_2} p(E_2) |A_{52}|^2
\ee
We assume that the characteristic energy of the operator $A$, which we call $\Delta$, is small compared to the spread in energies in the probability distribution $p(E_2)$. This will be correct if the probability distribution is canonical and $A$ is a simple operator with healthy UV behavior. (Similar techniques go through in the opposite regime, where the characteristic energy of the operator is large compared to the spread in energies, but we will not calculate this regime here.) In this case, we can treat $p(E_2)$ as a constant over the relevant range of energies, $p(E_2) \approx p(E_5)$ Note that
\be
\sum_{E_2} |A_{52}|^2 = \bra{E_5} A^\dagger A \ket{E_5}
\ee
so 
\be
\sum_{E_2} p(E_2) |A_{52}|^2 \approx p(E_5) \bra{E_5} A^\dagger A \ket{E_5}
\ee
Plugging this in gives for the quantum covariance
\be
\sum_{E_1, E_5} p(E_1) p(E_5) e^{i(E_5 - E_1)(t - t')}|A_{15}|^2 \bra{E_5} A^\dagger A \ket{E_5}
\ee
Using ETH once more, $ \bra{E_5} A^\dagger A \ket{E_5}$ depends only weakly on $\ket{E_5}$ so it can be taken out of the sum, giving
\be
\Bigl\llbracket \expval{A^\dagger A} \Bigr\rrbracket \sum_{E_1, E_5} p(E_1) p(E_5) e^{i(E_5 - E_1)(t - t')}|A_{15}|^2
\ee
For any time, we can rewrite this as a product of 2-point functions by defining a state in the doubled system
\be
\ket{T} \equiv {1 \over \sqrt{\sum_{E'} p(E')^2} }\sum_E p(E) \ket {E\ E}
\ee
Then the quantum covariance can be written as
\be
{\sum_{E'} p(E')^2} \Bigl\llbracket \expval{A^\dagger(0) A(0)} \Bigr\rrbracket \bra{T} A_L(t) A^\dagger_R(t') \ket{T}
\ee
This agrees with our result above. 

\section{Holographic Reconstruction of Gravitons}
In this chapter, we will apply the technique of holographic bulk reconstruction to reconstruct the metric fluctuation operator of a black hole. Specifically, we will consider a 5-dimensional AdS planar black hole, derive its reconstruction formula, and obtain the key coefficient expressions by solving the associated ordinary differential equations. Bulk reconstruction in AdS/CFT via the extrapolate dictionary was worked out in a beautiful series of papers beginning with the work of Hamilton, Kabat, Lifschytz, and Lowe (HKLL) \cite{Hamilton:2005ju, Hamilton:2006az}. Particular issues in reconstructing scalar fields near black holes were analyzed in \cite{Bousso:2012mh, Leichenauer:2013kaa, Morrison:2014jha}. In this paper we adapt the discussion to gravitational perturbations. Our results are simply an application of the HKLL technique.

\subsection{Fluctuations of an AdS planar black hole}\label{ads_perturbation}\label{sec_5.1}
Now we consider a small perturbation $h_{\mu\nu}$ added to the background metric $g_{\mu\nu}^0$, a 5-dimensional AdS planar black hole. The background is given by the metric
\begin{equation}\label{5.2}
d s^2=\frac{r^2}{R^2}\left(-f(r) d t^2+d x_1^2+d x_2^2+d x_3^2\right)+\frac{R^2}{r^2 f(r)} d r^2,
\end{equation}
where $R$ is a constant, and $f(r)=$ $1-r_0^4 / r^4$. The horizon is located at $r=r_0$, whose Hawking temperature is $T=r_0 / \pi R^2$. One can introduce a new coordinate $u=r_0^2 / r^2$ to rewrite the metric as
\begin{equation}\label{5.3}
d s^2=\frac{(\pi T R)^2}{u}\left(-f(u) d t^2+d x_1^2+d x_2^2+d x_3^2\right)+\frac{R^2}{4 u^2 f(u)} d u^2,
\end{equation}
where $f(u)=1-u^2$. Now the horizon is at $u=1$ in these coordinates, and we have a boundary at $u=0$. We will see that this coordinate system makes it easier to solve the equations for the gravitational perturbation later. For simplicity, we assume the perturbation to be independent on $x_1$-$x_2$ coordinates, and we choose the gauge where $h_{u\mu}=0$ for all $\mu$. This gauge is naturally self-consistent with the later holographic bulk reconstruction. The perturbed metric can be written as
\begin{equation}\label{perturbed_metric}
    g_{\mu\nu}(t,u,x_3) = g_{\mu\nu}^{0}(u) + h_{\mu\nu}(t,u,x_3).
\end{equation}
The translational invariance of the background metric makes us take the fluctuations to be of the form $h_{\mu\nu}(u)e^{-i\omega t + iqx_3}$. In this scenario, the gravitational perturbation can be categorized based on the rotational spin properties with respect to $\mathrm{O}(2)$ rotations within the $xy$ plane. More precisely, there exist three categories of perturbations (only the non-zero components of $h_{\mu \nu}$ are provided): \cite{Kovtun:2005ev}
\begin{itemize}
    \item Spin 0 (sound channel): $h_{tt}, h_{tx_3}, h_{x_3x_3}, h_{uu}, h_{tu}, h_{x_3u}$;
    \item Spin 1 (shear channel): $h_{tx_1}, h_{tx_2}, h_{x_1x_3}, h_{x_2x_3}, h_{ux_1}, h_{ux_2}$;
    \item Spin 2 (scalar channel): $h_{x_1x_2}$.
\end{itemize}
We will see later that the field equations of each symmetry channel are decoupled from each other because of the $\mathrm{O}(2)$ symmetry group. In fact, it is convenient to construct three gauge-invariants of the three classes respectively. Consider an infinitesimal change of coordinates, $x_{\mu}\rightarrow x_{\mu} - \xi_{\mu}$, where $\xi_{\mu}=\xi_{\mu}(u)e^{-i\omega t + iqx_3}$ with $\mu=t, x_1, x_2, x_3$. The metric changes by
\begin{equation}
    \delta g_{\mu\nu} = \nabla_{\mu} \xi_{\nu} + \nabla_{\nu} \xi_{\mu}.
\end{equation}
For the perturbed metric (\ref{perturbed_metric}), this can be seen as a transformation of the linearized field $h_{\mu\nu}$. At leading order in both $h_{\mu\nu}$ and $\xi_{\mu\nu}$, the field $h_{\mu\nu}$ is transformed as
\begin{equation}
    h_{\mu\nu} \rightarrow h_{\mu\nu} + \partial_{\mu}\xi_{\nu} + \partial_{\nu}\xi_{\mu} - g^{0 \alpha\sigma}( \partial_{\mu}g^0_{\sigma\nu} + \partial_{\nu}g^0_{\sigma\mu} - \partial_{\sigma}g_{\mu\nu}^0 )\xi_{\alpha}.
\end{equation}
If we take $\xi_{\mu}$ as gauge functions, we could build a gauge-invariant variable for each class of the linearized fields. These variables may be defined as
\begin{itemize}
    \item Shear channel: $Z_1=q H_{t x_1}+\omega H_{x_3 x_1}$;
    \item Sound channel: $Z_2=q^2 f H_{t t}+2 \omega q H_{t x_3}+\omega^2 H_{x_3 x_3}+q^2 f\left(1+\frac{a f^{\prime}}{a^{\prime} f}-\frac{\omega^2}{q^2 f}\right) H$;
    \item Scalar channel: $Z_3=H_{x_1 x_2}$,
\end{itemize}
where $H_{t t}=u h_{t t} / (\pi T R)^2 f, H_{t x_3}= u h_{t x_3} / (\pi T R)^2, H_{i j}=u h_{i j} / (\pi T R)^2 (i, j \neq t), H=u h /2(\pi T R)^2, a = (\pi TR)^2/u$.

From Einstein equations one can obtain three independent second-order ordinary differential equations satisfied by $Z_1$, $Z_2$, and $Z_3$. Before solving these equations in section \ref{slee}, we will first derive a holographic bulk reconstruction formula for these gauge invariants in the next section. 
\subsection{Reconstruction of the bulk gravitons}
The AdS/CFT extrapolate dictionary gives a simple relation between a local bulk operator near the boundary, and a local boundary operator. To see this we consider a bulk field $\Phi$ with normalized fall off close to the boundary of AdS. It is related to a boundary operator $O$ via
\begin{equation}
    \lim_{r\rightarrow \infty} r^{\Delta}\Phi(r,t,\Omega) = O(t,\Omega).
\end{equation}
But what is the CFT dual of a bulk operator $\Phi(r,t,\Omega)$ at finite $r$? In other words, we would like to find a mapping, where the bulk field far from the boundary is expressed in terms of boundary operators. This mapping formula is called holographic bulk reconstruction.
\par Now we consider the metric fluctuation, which can be written as a mode expansion
\begin{equation}\label{mode_expansion}
    h_{\mu\nu}(t,u,x_3)=\int \frac{d\omega dq}{(2\pi)^2}\left\{   a_{\omega,q}\phi(u;\omega,q)e^{-i\omega t+iqx_3}+a_{\omega,q}^{\dagger}\phi(u;\omega,q)^*e^{i\omega t -iqx_3} \right\},
\end{equation}
where $\phi(u;\omega,q)$ is a complete set of solutions to the field equation. In order to express this metric fluctuation with finite $u$ in terms of boundary data, we need to first find the extrapolate dictionary in this case. To do this, it is useful to rewrite the metric (\ref{5.3}) in Fefferman-Graham coordinates,
\begin{equation}
    ds^2 = \frac{R^2}{z^2}\left(  dz^2 + g_{\mu\nu}'(\mathbf{x}, z)dx^{\mu}dx^{\nu}  \right), 
\end{equation}
where the coordinate transformation between $z$ and $u$ is given by
\begin{equation}
    u=\frac{4 R^4 r_0^2 z^2}{4 R^8+r_0^4 z^4}.
\end{equation}
One can expand $g_{\mu\nu}'$ near the AdS boundary $z=0$,
\begin{equation}
z\rightarrow0:  \qquad  g_{\mu\nu}' = g_{\mu\nu}'^{(0)} + z^2 g_{\mu\nu}'^{(2)} + z^3 g_{\mu\nu}'^{(3)} + z^4 g_{\mu\nu}'^{(4)} + O(z^5).
\end{equation}
Then the metric coefficient is related to the boundary stress tensor by \cite{deHaro:2000vlm}
\begin{equation}\label{tg'}
    T_{\mu\nu} = \frac{R^3}{4\pi G_5} g_{\mu\nu}'^{(4)}
\end{equation}
And we have
\begin{equation}\label{gg'}
\begin{split}
    g_{\mu\nu} &= \frac{R^2}{z^2}g_{\mu\nu}' \\
               &= \frac{R^2}{z^2}g_{\mu\nu}'^{(0)}+R^2 g_{\mu\nu}'^{(2)} + z R^2 g_{\mu\nu}'^{(3)} + z^2 R^2 g_{\mu\nu}'^{(4)} + O(z^3).
\end{split}
\end{equation}
Substituting (\ref{perturbed_metric}) into (\ref{gg'}) and using (\ref{tg'}), we obtain that
\begin{equation}
    g_{\mu\nu}^{\scriptscriptstyle{0}\scriptstyle(2)} + h_{\mu\nu}^{(2)} = R^2 g_{\mu\nu}'^{(4)} = \frac{4\pi G_5}{R} T_{\mu\nu}.
\end{equation}
This is nothing but the extrapolate dictionary which relates the perturbed metric to the boundary stress tensor
\begin{equation}\label{extrapolate_dic}
    z\rightarrow0:  \qquad  h_{\mu\nu}^{(2)} = \frac{4\pi G_5}{R} T_{\mu\nu} - g_{\mu\nu}^{\scriptscriptstyle{0}\scriptstyle(2)},
\end{equation}
where the superscript $(1)$ denotes the first order coefficient of the expansion around $z=0$. Now let us come back to the mode expansion (\ref{mode_expansion}) and extract the first order coefficient near the boundary, which is given by
\begin{equation}\label{h1_mode_expansion}
    z\rightarrow0: \qquad h_{\mu\nu}^{(2)}(t,x_3) = \int \frac{d\omega dq}{(2\pi)^2}(a_{\omega,q} + a_{-\omega,-q}^{\dagger})\phi^{(2)}(\omega,q)e^{-i\omega t + iqx_3}.
\end{equation}
We can take the Fourier transform of (\ref{h1_mode_expansion}) to extract the creation and annihilation operators
\begin{equation}
    a_{\omega,q} + a_{-\omega,-q}^{\dagger} = \int dt dx\frac{h_{\mu\nu}^{\scriptscriptstyle (2)}(t,x_3)}{\phi^{\scriptscriptstyle (2)}(\omega,q)}e^{i\omega t - iqx_3}.
\end{equation}
Employing the extrapolate dictionary (\ref{extrapolate_dic}), we can write
\begin{equation}\label{aa}
    a_{\omega,q} + a_{-\omega,-q}^{\dagger} = \int dtdx \left\{  \frac{4\pi G_5}{R}T_{\mu\nu}(t,x_3) - g_{\mu\nu}^{\scriptscriptstyle{0}\scriptstyle(2)}   \right \}  \frac{e^{i\omega t - iqx_3}}{\phi^{\scriptscriptstyle (2)}(\omega,q)}.
\end{equation}
Then we insert (\ref{aa}) into the original mode expansion (\ref{mode_expansion}) to obtain the reconstruction formula
\begin{equation}
    \hat{h}_{\mu\nu}(t,\rho,x_3) = \int \frac{d\omega dq}{(2\pi)^2} \int dt'dx'\left\{ \frac{4\pi G_5}{R}\hat{T}_{\mu\nu}(t',x')  \right\} \frac{\phi(\rho,\omega,q)}{\phi^{\scriptscriptstyle (2)}(\omega,q)}e^{-i\omega(t-t')+iq(x_3-x')}.
\end{equation}
The formula can also be rewritten in a simpler form in Fourier space as
\begin{equation}
    \hat{h}_{\mu\nu}(u;\omega,q) = \frac{\phi(u,\omega,q)}{\phi^{\scriptscriptstyle (2)}(\omega,q)} \left( \frac{4\pi G_5}{R}\hat{T}_{\mu\nu}(\omega,q) - Q(\omega,q)  \right),
\end{equation}
where the function $Q(\omega,q)$ is defined as
\begin{equation}
    Q(\omega,q) = \int dt' dx' g_{\mu\nu}^{\scriptscriptstyle{0}\scriptstyle(2)}\frac{\phi(\rho;\omega,q)}{\phi^{\scriptscriptstyle (2)}(\omega,q)}e^{i\omega t' - iqx'}.
\end{equation}
Now we would like to find the bulk reconstruction formula for the gauge-invariants introduced in the previous section. Since these invariants are simple linear combinations of the metric perturbation in Fourier space, it is straightforward to write down
\begin{equation}\label{qd_final}
    \hat{Z}(u;\omega,q) = \frac{\phi(u;\omega,q)}{\phi^{\scriptscriptstyle (4)}(\omega,q)} \left(  \frac{4\pi G_5}{R} \hat{Z}^{\text{bdy}}(\omega,q) - Q(\omega,q)    \right),
\end{equation}
where we have introduced the boundary dual operators $\hat{Z}^{\text{bdy}}$ of these gauge invariants, which are identified with linear combinations of the Fourier transformed boundary stress tensors. Note that here the $\phi(\rho;\omega,q)$ now refer to the solutions of the ODEs satisfied by these gauge invariants $Z$ rather than the original components $h_{\mu\nu}$. 
\subsection{Solving linearized Einstein equations}\label{slee}
In this section, we shall solve the Einstein equation satisfied by the field perturbations. Recall that the perturbed metric reads
\begin{equation}
    g_{\mu\nu}(t,u,x) = g_{\mu\nu}^{0}(u) + h_{\mu\nu}(t,u,x).
\end{equation}
The contracted Ricci tensor derived from the perturbed form of the metric can be written as
\begin{equation}
    R_{\mu\nu} = R_{\mu\nu}^{0} + \delta R_{\mu\nu}.
\end{equation}
Here the expression of $\delta R_{\mu\nu}$ can be calculated by Palatini equation
\begin{equation}\label{delta_R}
    \delta R_{\mu\nu} = - \delta \Gamma^{\beta}_{\mu\nu;\beta} + \delta \Gamma^{\beta}_{\mu\beta;\nu},
\end{equation}
where the Christoffel symbol is given by
\begin{equation}\label{christoffel}
    \delta \Gamma^{\alpha}_{\beta\gamma}=\frac{1}{2} g^{\alpha\nu}\left( h_{\beta\nu;\gamma}+h_{\gamma\nu;\beta}-h_{\beta\gamma;\nu}    \right).
\end{equation}
Substituting (\ref{christoffel}) into (\ref{delta_R}), we obtain the final expression for $\delta R_{\mu\nu}$
\begin{equation}\label{delta_R_final}
    \delta R_{\mu\nu} = \frac{1}{2}g^{\alpha\sigma}
    \left(  -h_{\nu\sigma;\mu;\alpha}+h_{\mu\nu;\sigma;\alpha}+h_{\alpha\sigma;\mu;\nu}-h_{\mu\alpha;\nu;\sigma}    \right).
\end{equation}
The unperturbed 4-dimensional AdS planar black hole satisfies the equation
\begin{equation}
    R_{\mu\nu} = -\frac{4}{R^2} g_{\mu\nu},
\end{equation}
where $R$ denotes the AdS radius. Since the perturbed black hole should also be a solution to the Einstein equation, we can write
\begin{equation}\label{delta_R_h}
    \delta R_{\mu\nu} = -\frac{4}{R^2} h_{\mu\nu}.
\end{equation}
Combining (\ref{delta_R_final}) with (\ref{delta_R_h}), we get the field equations for the $h_{\mu\nu}$. As we discussed in section (\ref{sec_5.1}), these equations can be divided into three classes, in each of which they form an ordinary differential equation satisfied by the corresponding gauge-invariant. We first consider the simplest class, the scalar channel, where the field equation is found to be
\begin{equation}
    Z_3^{\prime \prime}-\frac{1+u^2}{u f} Z_3^{\prime}+\frac{\tilde{\omega}^2-\tilde{q}^2 f}{u f^2} Z_3=0,
\end{equation}
where we have defined 
\begin{equation}
    \tilde{\omega}=\frac{\omega}{2 \pi T}, \quad \tilde{q}=\frac{q}{2 \pi T}.
\end{equation}
To keep it concise, we have omitted the detailed derivation process of this equation in the main body of the text. For the shear channel, we can write down two field equations
\begin{gather}
    \begin{aligned}
& H_{x_3 x_1}^{\prime}=-\frac{\tilde{\omega}}{\tilde{q} f} H_{t x_1}^{\prime}, \\
& u H_{t x_1}^{\prime \prime}= H_{t x_1}^{\prime}+\frac{\tilde{\omega} \tilde{q}}{ f} H_{x_3 x_1}+\frac{\tilde{q}^2}{ f} H_{t x_1},
\end{aligned}
\end{gather}
We can then obtain the ODE for $Z_1$ by manipulating these two equations, which is given by
\begin{equation}\label{shear_eq}
   Z_1^{\prime \prime}+\frac{\left(\tilde{\omega}^2-\tilde{q}^2 f\right) f-u \tilde{\omega}^2 f^{\prime}}{u f\left(\tilde{q}^2 f-\tilde{\omega}^2\right)} Z_1^{\prime}+\frac{\tilde{\omega}^2-\tilde{q}^2 f}{u f^2} Z_1=0 .
\end{equation}
Similarly, the field equations for the sound channel, $h_{tt}$, $h_{tx_3}$ and $h_{x_3x_3}\neq0$ lead to an ODE for $Z_2$
\begin{equation}
\begin{aligned}
Z_2^{\prime \prime} & -\frac{3 \tilde{\omega}^2\left(1+u^2\right)+\tilde{q}^2\left(2 u^2-3 u^4-3\right)}{u f\left(3 \tilde{\omega}^2+\tilde{q}^2\left(u^2-3\right)\right)} Z_2^{\prime} \\
& +\frac{3 \tilde{\omega}^4+\tilde{q}^4\left(3-4 u^2+u^4\right)+\tilde{q}^2\left(4 u^5-4 u^3+4 u^2 \tilde{\omega}^2-6 \tilde{\omega}^2\right)}{u f^2\left(3 \tilde{\omega}^2+\tilde{q}^2\left(u^2-3\right)\right)} Z_2=0
\end{aligned}
\end{equation}
We need to find the solutions $\phi(u)$ for these equations and substitute them into (\ref{qd_final}). The complexity of the equations results in the absence of exact solutions. Therefore, we need to consider employing corresponding approximation methods to solve them within different parameter ranges of $\tilde{\omega}$ and $\tilde{q}$. Here we will solve the equations for the scalar channel in a specific parameter regime where $\tilde{q}\gg 1$, $\tilde{\omega}\sim1$ .
\subsection{$\tilde{q}\gg 1$, $\tilde{\omega}\sim1$}
In this parameter range, the hydrodynamic limit is not applicable, and we cannot solve these equations using perturbative expansions. In such cases, the WKB approximation is a commonly used method to handle these second-order equations. Below, we will use the WKB approximation to solve a simplest case, the equation for the scalar channel. 
We need to solve an equation that can be written in the following form:
\begin{equation}
Z_3^{\prime \prime}-P(u) Z_3^{\prime}+Q(u) Z_3=0
\end{equation}
where $P(u)=-\frac{1+u^2}{u f}$, $Q(u)=\frac{\tilde{\omega}^2-\tilde{q}^2 f}{u f^2}$, and $f(u)=1-u^2$ . To facilitate the use of WKB approximation, we need to rewrite the equation in standard form. To do so, one can introduce a function \( \Omega(u) \) such that \( Z_3(u) = \Omega(u)Y(u) \), where \( Y(u) \) is a newly introduced variable that satisfies the following equation:
\begin{equation}\label{schordinger_like_eq}
    -Y^{\prime \prime}+V(u) Y=0,
\end{equation}
where $V(u)$ can be written as:
\begin{equation}
V(u)=-Q(u)+\frac{P^{\prime}(u)}{2}+\frac{P^2(u)}{4}
\end{equation}.

\begin{figure}[h]
    \centering
    \includegraphics[width=0.8\textwidth]{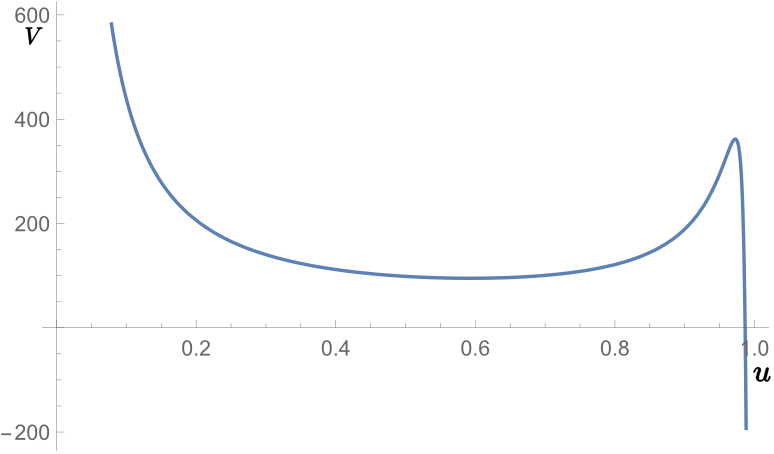}
    \caption{The graph of the potential V(u). The intersection point of $V$ and the x-axis is denoted as $u_0$.}
    \label{fig1}
\end{figure}

For this to work, the function \( \Omega(u) \) needs to take the following form:
\begin{equation}
\Omega(u)=\exp \left\{-\frac{1}{2} \int_{u_0}^u p\left(u^{\prime}\right) d u^{\prime}\right\}.
\end{equation}
The equation (\ref{schordinger_like_eq}) resembles a Schrodinger equation, where $V(u)$ can be viewed as a potential. The functional curve of $V(u)$  is plotted in Figure \ref{fig1} for the parameter range we are considering ($\tilde{q}\gg 1$, $\tilde{\omega}\sim1$). We can divide this function into multiple intervals and apply the WKB approximation to calculate the local solutions in each region. Then, we can stitch them together using boundary conditions to form a global solution.

To ensure the effectiveness of the WKB approximation, one has to require the following conditions:
\begin{equation}
\left|k^{\prime}(u)\right| \ll|k(u)|^2,
\end{equation}
where $k$ denotes the momentum. This is equivalent to
\begin{equation}
V^{\prime}(u) \ll V^{3 / 2}(u).
\end{equation}
Substituting the approximate expression for $V(u)$ at large $q$, one can easily obtain:
\begin{equation}\label{wkb_condition}
\tilde{q}^2 u \gg 1.
\end{equation}
We need to pay attention to this condition when solving for the local solutions in each interval and utilize it in possible approximations.

The first interval is at the left boundary, where $u\rightarrow0$. The potential there can be approximated as  
\begin{equation}
V(u) \approx \frac{3}{4 u^2}-\frac{\tilde{\omega}^2-\tilde{q}^2}{u}.
\end{equation}
Note that the equation for this potential is solvable, and its solution is composed of a linear combination of modified Bessel functions, which is given by
\begin{equation}\label{solution_y}
\begin{aligned}
Y(u) & =\frac{i \sqrt{u}\left(4 \tilde{q}^2-4 \tilde{\omega}^2\right)^{3 / 2}}{4\left(\tilde{q}^2-\tilde{\omega}^2\right)} I_2\left(  2\sqrt{\left(\tilde{q}^2-\tilde{\omega}^2\right) u} \right)  B_1 \\
& +2\sqrt{\left(\tilde{q}^2-\tilde{\omega}^2\right) u} K_2\left(2\sqrt{\left(\tilde{q}^2-\tilde{\omega}^2\right) u}\right) B_2,
\end{aligned}
\end{equation}
where $B_1$ and $B_2$ are arbitrary constants.

The condition (\ref{wkb_condition}) makes us introduce a variable $x=2 \sqrt{\left(\tilde{q}^2-\tilde{\omega}^2\right) u}$ . The asymptotic form for $I_{\nu}(x)$ when $x\gg 1$ is given by $I_\nu(x) \approx e^x / \sqrt{2 \pi x}$. And the asymptotic solution for $K_{\nu}(x)$ when $x\gg1$ reads
\begin{equation}
K_\nu(x) \approx \sqrt{\frac{\pi}{2 x}} e^{-x}\left[1+\frac{4 \nu^2-1}{8 x}+\frac{\left(4\nu^2-1\right)\left(4 \nu^2-9\right)}{128 x^2}+\ldots\right].
\end{equation}
Therefore, one can find that the second term in (\ref{solution_y}) can be written down as
\begin{equation}
    x K_2(x)=\frac{1}{2}x\Gamma(2)(\frac{2}{x})^2=\frac{2}{x},
\end{equation}
which vanishes when x is very large. Then one can only keep the first term to write down the asymptotic solution for $Y(u)$ at $u\rightarrow0$:
\begin{equation}\label{yto0}
    Y(u)=\frac{i}{\sqrt{2\pi}}e^x\sqrt{x}=\frac{i}{\sqrt{\pi}}\left(\tilde{q}^2-\tilde{\omega}^2\right)^{1 / 4} u^{1 / 4} e^{2 \sqrt{\left(\tilde{q}^2-\tilde{\omega}^2\right) u}}.
\end{equation}

The second interval we are considering is neither close to $u\rightarrow0$ nor far from $u=1$; it's an intermediate region. In this area, there is a turning point where the potential curve intersects with the y-axis. Near this turning point, the WKB approximation usually breaks down. However, we can use classical methods to stitch together the equations on both sides of the turning point. The effect of this is manifested as a phase difference in the equations on either side.

The WKB ansatz for this region is given by this standard form:
\begin{equation}
\begin{aligned}
Y(u)= & C_1 V(u)^{-1/4} \exp \left\{-\int_{u_0}^{u} \sqrt{V\left(u^{\prime}\right)} d u^{\prime}\right\} \\
& +C_2 V(u)^{-1 / 4} \exp \left\{+\int_{u_0}^u \sqrt{V\left(u^{\prime}\right)} d u^{\prime}\right\}.
\end{aligned}
\end{equation}
Here $u_0$ denotes the turning point where $V(u_0)=0$. Consider $\tilde{q}\gg1$, integrating the exponent, one can obtain the following expression as $u\rightarrow0$
\begin{equation}
\begin{aligned}
Y(u)&=C_1 \left(\tilde{q}^2 -\tilde{\omega}^2  \right)^{-1 / 4} u^{1 / 4}  e^{-2 \sqrt{\left(\tilde{q}^2-\tilde{\omega}^2\right) u}+D\tilde{q}} \\
& +C_2\left(\tilde{q}^2 -\tilde{\omega}^2  \right)^{-1 / 4} u^{1 / 4} e^{2 \sqrt{\left(\tilde{q}^2-\tilde{\omega}^2\right) u}-D\tilde{q}},
\end{aligned}
\end{equation}
where $D$ is a constant given by
\begin{equation}
    D = 2\sqrt{u_0}\times{}_2F_1\left(\frac{1}{4}, \frac{1}{2}, \frac{5}{4}, u_0^2\right),
\end{equation}
where ${}_2F_1$ denotes the hypergeometric function. This is an $O(1)$ constant as $u_0$ goes to zero. One can determine the values of the arbitrary constants $C_1$
and $C_2$ through boundary conditions. When $u\rightarrow0$, the expression for $Y(u)$ will tend to become the equation (\ref{yto0}) , which will immediately make $C_1=0$, and by comparing with (\ref{yto0}), one can find the value of $C_2$. Ultimately, the constant values we obtain are:
\begin{equation}
\left\{\begin{array}{l}
C_1=0 \\
C_2=\frac{i}{\sqrt{\pi}}\left(\tilde{q}^2-\tilde{\omega}^2\right)^{1 / 2} e^{D\tilde{q}}.
\end{array}\right.
\end{equation}
Up to this point, we have obtained the WKB solution for the interval at $u<u_0$. The WKB solution for $u>u_0$ can be bridged across the turning point $u_0$ using traditional patching methods. Therefore, the WKB solution in the entire  region takes the following form:
\begin{equation}\label{Y_final}
Y(u)=\left\{\begin{array}{l}
\frac{i}{\sqrt{\pi}}\left(\tilde{q}^2-\tilde{\omega}^2\right)^{1 / 2} e^{D\tilde{q} } V(u)^{-1 / 4}\exp \left(\int_{u_0}^u \sqrt{V\left(u^{\prime}\right)} d u^{\prime}\right) \quad u<u_0 \\
\frac{2 i}{\sqrt{\pi}}\left(\tilde{q}^2-\tilde{\omega}^2\right)^{1 / 2}e^ {D\tilde{q}} V(u)^{-1 / 4} \sin \left(\int_{u_0}^u \sqrt{\left|V\left(u^{\prime}\right)\right|} d u^{\prime}+\frac{\pi}{4}\right) \quad u>u_0
\end{array}\right.
\end{equation}
We have now obtained the WKB solution $Y(u)$ for the entire interval $0<u<1$. The next step is to find the corresponding solutions for the gauge variable $Z_3(u)$ through $\Omega(u)$. The expression for $\Omega(u)$ can be obtained through an integral:
\begin{equation}\label{omega_final}
\begin{aligned}
\Omega(u) & =\exp \left\{-\frac{1}{2} \int_{u_1}^u P\left(u^{\prime}\right) d u^{\prime}\right\} \\
& =\left[\left(\frac{1-u^2}{u}\right) / \frac{1-u_1^2}{u_1}\right]^{-1 / 2}\\
& =A(\frac{1-u^2}{u})^{-1/2},
\end{aligned}
\end{equation}
where $A$ is an arbitrary constant that will be canceled off in the following calculation. Note that in equation (\ref{qd_final}), we are only concerned with the behavior of the solution $\phi(u)$ in the segments 
$u\rightarrow0$ and $u\rightarrow1$ . Therefore, one has to look for the approximate solutions of $\phi(u)$ on these two sides. It is easy to see that $Y(u)$ at $u\rightarrow0$ contains only the first term, the modified Bessel function in which is given by
\begin{equation}
\begin{aligned}
I_2\left(\sqrt{2 (\tilde{q}^2- \tilde{\omega}^2)u}\right) & =\frac{1}{\Gamma(3)}\left(\tilde{q}^2-\tilde{\omega}^2\right) u  +\frac{1}{\Gamma(4)}\left(\tilde{q}^2-\tilde{\omega}^2\right)^2 u^2+O\left(u^4\right),
\end{aligned}
\end{equation}
and thus the function $Y(u)$ can be expanded as
\begin{equation}\label{final_Y}
\begin{aligned}
Y(u)= & \frac{i \sqrt{u}\left(4 \tilde{q}^2-4 \tilde{\omega}^2\right)^{3 / 2}}{4\left(\tilde{q}^2-\tilde{\omega}^2\right)} I_2\left(\sqrt{2 (\tilde{q}^2- \tilde{\omega}^2)u}\right) \\
= & 2 i\left(\tilde{q}^2-\tilde{\omega}^2\right)^{1 / 2}\left\{\begin{array}{c}
\frac{1}{2}\left(\tilde{q}^2-\tilde{\omega}^2\right) u^{3 / 2}+\frac{1}{6}\left(\tilde{q}^2-\tilde{\omega}^2\right)^2 u^{5 / 2} 
\left.+O\left(u^4\right)\right\}.
\end{array}\right.
\end{aligned}
\end{equation}
Combining  (\ref{omega_final}) with (\ref{final_Y}) , one could get the solution $\phi(u)$ to $Z_3(u)$ as $u\rightarrow 0$,
\begin{equation}
u \rightarrow 0: \quad \phi(u)=i A\left(\tilde{q}^2-\tilde{\omega}^2\right)^{3 / 2} u^2+O\left(u^3\right)
\end{equation}
Since $\phi^{\scriptscriptstyle (4)}$ is the expansion of $\phi$ around $z=0$, one can read off the value of $\phi^{\scriptscriptstyle (4)}$ based on the coordinate relation between $z$ and $u$ near $u=0$, which is given by
\begin{equation}
    u=\frac{r_0^2}{R^4}z^2.
\end{equation}
The expression of $\phi^{\scriptscriptstyle (4)}$ is then given by
\begin{equation}\label{phi_4}
    \phi^{\scriptscriptstyle (4)}=A\left(\tilde{q}^2-\tilde{\omega}^2\right)^{3/2}\left( \frac{r_0^2}{R^4}\right)^2
\end{equation}
So far we have obtained the denominator of the ratio appearing in the equation (\ref{qd_final}), and we will consider the numerator $\phi(u)$ now. It's worth noting that we are particularly interested in metric perturbations near the black hole horizon. Therefore, we will let u approach $1$ and estimate the behavior of 
$\phi(u)$ in this vicinity. Putting (\ref{Y_final}) and (\ref{omega_final}) together, one gets the following expression
\begin{equation}\label{phi_1}
\begin{aligned}
u\rightarrow1: \quad \phi(u)&=Au^{1/2}e^{D\tilde{q}}\left(\tilde{q}^2-\tilde{\omega}^2\right)^{1 / 2} \left(1+\tilde{\omega}^2\right)^{-1 / 4} \\
&\times\sin \left(\int_{u_0}^u \sqrt{\left|V\left(u^{\prime}\right)\right|} d u^{\prime}+\frac{\pi}{4}\right) 
\end{aligned}
\end{equation}
Note that the constant arising from the integral of $\Omega(u)$ appear in both the expression of the numerator and the denominator, which will be canceled out in physically meaningful expressions, as only the ratio will appear within them.  Now, we have successfully solved the perturbation equations for the metric field in the context of an $\text{AdS}_5$ planar black hole and obtained its approximate form in the region of interest to us. These are essential elements for calculating the quantum deviation of the metric, which will be demonstrate in the following sections. We will present formulas for calculating quantum deviation and discuss how to use the elements solved previously to substitute into these formulas, thereby obtaining specific expressions for quantum deviation.

\section{The Quantum Deviation of an AdS Planar Black Hole}
\label{sec-bh}
\subsection{The formula for an AdS planar black hole}
In the previous section we have constructed several gauge-invariant variables of the metric fluctuation and expressed them in terms of the corresponding boundary dual operators. Now we apply the formula proposed in section \ref{roqd} to calculate the quantum deviation of the fluctuation of the metric. Here we focus on a canonical ensemble where the inverse temperature is set to be $2\beta$. Recall that in this case the quantum deviation of an operator $O$ is calculated by a connected two-point function in a thermofield double state \cite{Freivogel:2021ivu}
\begin{equation}
        \Delta_O^2 = \frac{Z(2\beta)}{Z(\beta)^2}\Bigl( \bra{\text{TFD}_{2\beta}}O_L O_R\ket{\text{TFD}_{2\beta}}_c + O(e^{-S}) \Bigr).
\end{equation}
And the operator $O$ should be set as the gauge-invariant $Z$, which can be expressed in terms of boundary data as
\begin{equation}
    \hat{Z}(u;\omega,q) = \frac{\phi(u;\omega,q)}{\phi^{\scriptscriptstyle (4)}(\omega,q)} \left(  \frac{4\pi G_5}{R} \hat{Z}^{\text{bdy}}(\omega,q) - Q(\omega,q)    \right).
\end{equation}
Note that the left hand side of this formula refers to the bulk operator while the right hand side contains an operator from the boundary side. There is also a function $Q(\omega,q)$ on the right hand side, which has no contribution to the quantum deviation. Actually, this function will vanish in the final result if put inside a connected two-point correlator. The quantum deviation of $\hat{Z}(u;\omega,q)$ can be expressed in this form
\begin{equation}\label{delta_z_final}
\begin{split}
    \Delta_{\hat{Z}}^2 &=\frac{Z_{2\beta}}{Z_{\beta}^2}(\frac{4\pi G_5}{R})^2 \left(\frac{\phi(u,\omega,q)}{\phi^{\scriptscriptstyle (4)}(\omega,q)}\right)^2\\
    &\times \left\{ \bra{\text{TFD}_{2\beta}} \hat{Z}_L^{\text{bdy}}(\omega,q)  \hat{Z}_R^{\text{bdy}}(\omega,q) \ket{\text{TFD}_{2\beta}} -    \langle \hat{Z}^{\text{bdy}}(\omega,q)\rangle^2_{2\beta}   \right\}.
\end{split}
\end{equation}
The operators $Z_L$ and $Z_R$ within the two-point correlator are put on the left and right side of the TFD state respectively. And the one-point function within the bracket is a thermal correlator with temperature $2\beta$. Thus we have obtained the expression of the quantum deviation of the metric fluctuation for a 5-dimensional planar AdS black hole.

Note that this formula mainly consists of two parts. The first part is the ratio of the solution of the gauge variable's equation that we need to compute, in the region and with the boundary values that concern us. The expression for this in AdS has already been provided in the previous section. The second part is the connected correlation function of the gauge variable operator at the boundary. This correlation function is defined on the thermofield double state, which is dual to the 5-dimensional AdS planar black hole. In the sections that follow, we will compute this correlation function.

\subsection{The two-point thermal correlator}

In this section we would like to calculate the two-point function within (\ref{delta_z_final}). We can compute this correlator holographically. Considering a state dual to an eternal black hole geometry, the correlator can be calculated by turning on sources at the boundary of the black hole geometry. 
\begin{figure}[h]
    \centering
    \includegraphics[width=0.4\textwidth]{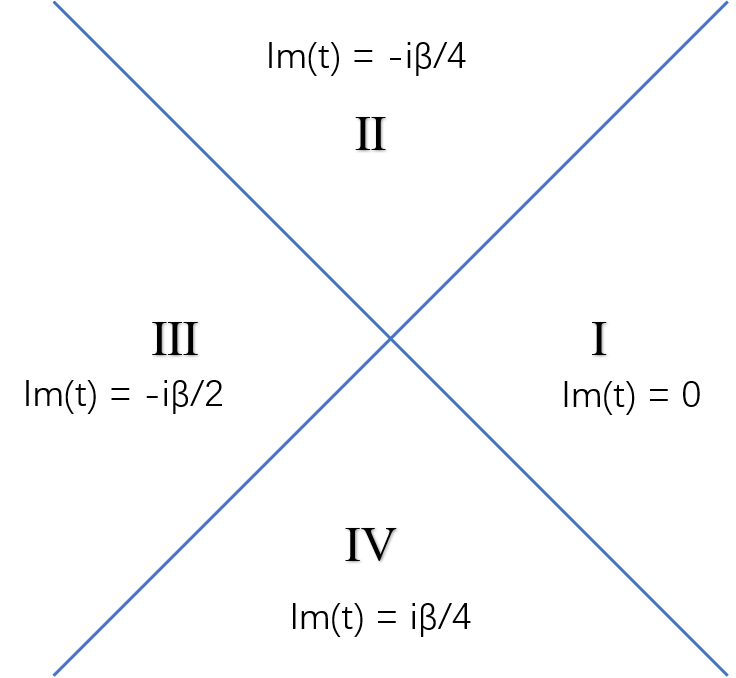}
    \caption{Complexified coordinates for the AdS black hole.}
    \label{fig2}
\end{figure}
In order to describe both the left and right boundary of the eternal black hole, one can employ Kruskal coordinate $(T,X)$ covering the global structure of the maximally extended spacetime. Nevertheless, it is more convenient for us to use four Schwarzchild patches to discuss this global extension. As shown in Figure \ref{fig2}, the four patches include two asymptotic regions (\Rmnum{1} and \Rmnum{3}), as well as the region inside the horizon (\Rmnum{2}) and the white hole (\Rmnum{4}).
\par We can embed the four patches in complexified Schwarzchild time,
\begin{equation}
    t = t_L + it_E,
\end{equation}
where $t_L$ and $t_E$ denote Lorentzian and Euclidean time slices respectively. For each patch, the imaginary part $t_E$ is a constant. We set $t_E$ to be zero in the right asymptotic region. Supposing the inverse temperature is $\beta$ , every time we cross a horizon, the imaginary part will be shifted by $i\beta/4$. If we just keep the imaginary part $t=it_E$, we can describe an Euclidean AdS black hole with a periodic time $t_E$, and the period of $t_E$ is $\beta$. Note that region \Rmnum{1} and region \Rmnum{3} have opposite directions in Lorentz time. Therefore, we can move a point from the boundary of region \Rmnum{1} to the symmetric point on the boundary of region \Rmnum{3} by transforming $t \rightarrow -t - i\beta/2$. This analytical continuation trick is useful when we compute a two-point function of operators inserted on the left and right boundary respectively. We focus on the two-point correlator in (\ref{delta_z_final}), which can be Fourier transformed as
\begin{equation}
\begin{split}
    &\qquad \qquad\bra{\text{TFD}_{2\beta}} \hat{Z}_L^{\text{bdy}}(\omega,q)  \hat{Z}_R^{\text{bdy}}(\omega,q) \ket{\text{TFD}_{2\beta}}\\
    &=\int dt dx \int dt' dx' \bra{\text{TFD}_{2\beta}} \hat{Z}_L^{\text{bdy}}(t,x)  \hat{Z}_R^{\text{bdy}}(t',x') \ket{\text{TFD}_{2\beta}} e^{-i\omega(t+t')+iq(x+x')}.
\end{split}
\end{equation}
Since the inverse temperature of the black hole is $2\beta$, we can use complexified Schwarzchild time as discussed above to replace $\hat{Z}_R^{\text{bdy}}(t',x')$ with $\hat{Z}_R^{\text{bdy}}(-t'-i\beta,x')$ in the Fourier integral. This replacement leads to a simple relation
\begin{equation}
\begin{split}
    &\qquad \bra{\text{TFD}_{2\beta}} \hat{Z}_L^{\text{bdy}}(\omega,q)  \hat{Z}_R^{\text{bdy}}(\omega,q) \ket{\text{TFD}_{2\beta}}\\
    &= e^{-\omega\beta} \bra{\text{TFD}_{2\beta}} \hat{Z}_L^{\text{bdy}}(\omega,q)  \hat{Z}_L^{\text{bdy}}(-\omega,q) \ket{\text{TFD}_{2\beta}}\\
    &=e^{-\omega\beta} \langle \hat{Z}^{\text{bdy}}(\omega,q)  \hat{Z}^{\text{bdy}}(-\omega,q)  \rangle_{2\beta}
\end{split}
\end{equation}
Therefore we have rewritten the two-sided correlator as an ordinary thermal correlator on the boundary of an AdS planar black hole, which is easier to compute via holography.  

Next, we will follow the idea presented in Kovtun's paper to calculate this thermal correlator. First, we need to identify the relevant boundary gravitational action, which in our notation can be written as:
\begin{equation}
S=-\frac{\pi^2 N_c^2 T^4}{8} \lim _{u \rightarrow 0} \int \frac{d \omega d q}{(2 \pi)^2} \frac{f(u)}{u} \phi_3^{\prime}(u, \omega,q) \phi_3(u,-\omega,-q)
\end{equation}
Proceeding as in \cite{Kovtun:2005ev}, one could express the derivatives of the fields in terms of the boundary values of the fields to obtain that
\begin{equation}\label{correlator_zz}
\left\langle\hat{Z}_3^{\text{bdy}}(\omega, q) \hat{Z}_3^{\text{bdy}}(-\omega, q)\right\rangle=\frac{-\delta S^2}{\delta \phi_3^0(\omega, q) \delta \phi_3^{0}(-\omega, q)}=\frac{\pi^2 N_c^2 T^4 \mathcal{B}_{(3)}}{2 \mathcal{A}_{(3)}},
\end{equation}
where $\mathcal{A}_{(3)}$ and $\mathcal{B}_{(3)}$ are the coefficients from the solution to the boundary asymptotic equation for $\phi_3(u)$. When we consider the parameter region where $\tilde{q}\gg1$, these coefficients can be read off from the equations obtained in the above section
\begin{equation}
\left\{\begin{array}{l}
\mathcal{B}_{(3)}=\frac{i\left(\tilde{q}^2-\tilde{\omega}^2\right)^{3 / 2}}{A} \\
\mathcal{A}_{(3)}=\frac{\left(\tilde{q}^2-\tilde{\omega}^2\right)^{-1 / 2}}{A}
\end{array}\right.
\end{equation}
Putting all of these elements together, one can write down the expression for the two-point correlator on the TFD state
\begin{equation}\label{corr}
\bra{\text{TFD}_{2\beta}} \hat{Z}_L^{\text{bdy}}(\omega,q)  \hat{Z}_R^{\text{bdy}}(\omega,q) \ket{\text{TFD}_{2\beta}}
=\frac{1}{2}i\pi^2N_c^2 T^4 e^{-\omega \beta}\left(\tilde{q}^2-\tilde{\omega}^2\right)^2.
\end{equation}
Note that this correlator is valid only when $\tilde{q}\gg1$ because we have used the asymptotic solution in this region to obtain the coefficients in (\ref{correlator_zz}).
\subsection{Results and discussions}
Now, having completed the preparatory work, we are ready to calculate the quantum deviation of the gauge variables of interest. We focus on the example of a 5-dimensional AdS planar black hole. Recall the formula for calculating quantum deviation (\ref{delta_z_final}). 
\begin{equation}\label{4.11}
\begin{split}
    \Delta_{\hat{Z}}^2 &=\frac{Z_{2\beta}}{Z_{\beta}^2}(\frac{4\pi G_5}{R})^2 \left(\frac{\phi(\rho,\omega,q)}{\phi^{\scriptscriptstyle (4)}(\omega,q)}\right)^2\\
    &\times \left\{ \bra{\text{TFD}_{2\beta}} \hat{Z}_L^{\text{bdy}}(\omega,q)  \hat{Z}_R^{\text{bdy}}(\omega,q) \ket{\text{TFD}_{2\beta}} -    \langle \hat{Z}^{\text{bdy}}(\omega,q)\rangle^2_{2\beta}   \right\}.
\end{split}    
\end{equation}
Note that the one-point correlator in the formula should actually vanish, as the operator $Z$ is a linear combination of components of metric perturbations, whose expectation values for these fluctuations are zero. Therefore, we only need to substitute the expressions for the two-point correlators and the ratio originating from the reconstruction formula into the equation to obtain the expression for the quantum deviation.

Let's consider the case where $\tilde{q}\gg 1$, $\tilde{\omega}\sim1$. This is the parameter regime of greatest interest to us. Plugging  (\ref{phi_4}), (\ref{phi_1}),  and (\ref{corr}) into (\ref{4.11}), we can derive the following expression for the quantum deviation in the scalar channel for a 5-dimensional AdS planar black hole:
\begin{equation}
\Delta_{\hat{Z}_{(u)}}^2 \sim u \frac{Z_{2 \beta}}{Z_\beta^2}\left(\frac{R}{\beta}\right)^2 e^{-4 \pi \tilde{\omega}+2 D\tilde{q}}\left(1+\tilde{\omega}^2\right)^{-1 / 2} \quad \quad \tilde{q} \gg 1, \tilde{\omega}\sim1
\end{equation}
Note that this derivation is proportional to $u$, the radial coordinate. Here $u_0$ denotes the position where the WKB potential vanishes. The inverse temperature of the black hole is denoted by $\beta$ , and $R$ is the AdS radius. Since we would like to investigate the region near the black hole horizon, we simply let $u$ goes to $1$. 

Now we can discuss some constraints on the order of magnitude of this quantity. Firstly, we notice that the ratio of the partition function $Z_{2\beta}/Z_{\beta}^2$ should be estimated to be $e^{-\#S}$  , where $\#$ is a number of order 1 and $S$ denotes the entropy of the black hole. The presence of this quantity makes the whole expression tend to be very small. If we assume that the entire expression is of $O(1)$ and considering the fact that $D$ is an $O(1)$ constant, then it necessitates $\tilde{q}$ in the exponent to be very large, reaching the order of magnitude of $S$. In other words, only the modes with a wave number at the order of magnitude of $S$ will have a quantum deviation at the detectable $O(1)$ scale. We need to conduct a physical existence analysis for such modes. Recall that we are working on the metric of a 5-dimensional black brane, 
\begin{equation}
d s^2=\frac{(\pi T R)^2}{u}\left(-f(u) d t^2+d x_1^2+d x_2^2+d x_3^2\right)+\frac{R^2}{4 u^2 f(u)} d u^2.
\end{equation}
The coordinate wavelength of the mode is thus given by
\begin{equation}
    \lambda_{\text{coord}} = \frac{2\pi}{q}=\frac{\beta}{\tilde{q}}.
\end{equation}
The relation between the coordinate wave length and the physical wavelength in this geometry is given by
\begin{equation}
    \lambda_{\text{phys}} = \frac{\pi T R}{u^{1/2}}\lambda_{\text{coord}}.
\end{equation}
When we specialize to the modes with $\tilde{q}$ of order $S$, it is easy to see that the physical wavelength near the horizon takes the following form
\begin{equation}
    \lambda_{\text{phys}}\sim\frac{R}{S}\sim \frac{R}{A} l_p^3,
\end{equation}
where $A$  denotes the area of the black hole, and $l_p$ is the Planck length. Here we have used the fact that the black hole entropy $S\sim A/l_p^3$. Then using $A\sim R^3$, we can compare this wavelength with the Planck length
\begin{equation}
    \frac{\lambda_{\text{phys}} }{l_p}\sim\frac{l_p^2}{R^2}.
\end{equation}
Clearly, this ratio indicates that the physical wavelength of this mode is much smaller than the Planck scale. Since we adopted a semi-classical approximation in our derivation, this approximation is no longer valid at such a scale. Therefore, we are compelled to conclude that under the assumption that the semi-classical approximation is valid, our previous assumption that $\tilde{q}\sim S$ is incorrect. In other words, for the quantum deviation of the channel we are considering, it cannot be at the $O(1)$ scale.

More generally, if we impose that the physical wavelength is longer than the Planck scale, the above equations imply
\be
\tilde q \lesssim {R \over l_P}
\ee
so $\tilde q \ll S$ in the semiclassical approximation.

\paragraph{Higher point functions.}
Finally, we can consider the quantum deviation of higher point functions. The bulk-boundary mapping relates bulk operators to boundary operators, so the quantum deviation of the $n$-point function requires $n$ copies of the bulk-boundary mapping. In the same near horizon, large $\tilde q$ limit considered above, this gives
\be
\Delta^2_{g^n} \sim e^{D\tilde q n} e^{-S}
\ee
where we have only kept track of the exponential dependence. Recall that $D \approx 2.6$ is the constant computed above.

This formula indicates that low-point functions are insensitive to the microstate, while high-point functions become more sensitive. As we commented in the introduction, we expect that corrections to the bulk-boundary mapping become important for high-point functions; it would be interesting to determine the

\newpage
\section{Discussion}
In this paper, we have considered the estimation of the magnitude of metric fluctuations for black holes. We introduced the concept of quantum deviation, reviewed computational techniques for black hole perturbation modes, and outlined the steps for holographic bulk reconstruction. Utilizing these techniques, we derived a general formula for the metric of a 5-dimensional AdS black hole, which can in principle be modified to generalize to other dimensions. When considering a 5-dimensional black brane geometry, we computed the quantum deviation of its perturbation modes in the large $\tilde{q}$ regime and placed constraints on its magnitude. 

Numerous aspects of this subject remain unexplored and and they serve as future directions of the research:

\textbf{Generalize the result to more dimensions and more parameter regimes.}  We are currently focusing on a 5-dimensional AdS spacetime, but it is worth considering extending this to higher-dimensional spacetimes, particularly four dimensions.  Additionally, we could also broaden our examination of modes with different wave numbers; although modes with large $\tilde{q}$ are of most physical interest, other parameter regimes are also worthy of study.

\textbf{Identify all possible degrees of freedom
of the black hole perturbation. }Among the various components of the black hole metric, we have only concentrated on the simplest scalar channel for quantum deviation, which does not include the full degrees of freedom of gravity. To make this story complete, it is necessary to identify every channel of the perturbation and estimate the corresponding quantum deviation.

\textbf{State Dependence.} As we commented in the introduction, our results rely on the assumption that bulk reconstruction is state-independent for small perturbations outside the horizon. If one wants to evade our conclusions, it would be interesting to exploit this loophole in more detail, and to develop the necessary corrections to the bulk-boundary dictionary.


\textbf{Beyond AdS/CFT.}  The calculations we've conducted are within the framework of the AdS/CFT duality. Utilizing holographic bulk reconstruction, we are able to map the black hole perturbation operator to the boundary, as our formula for calculating quantum deviation is formulated in the boundary field theory. However, if we wish to consider more realistic scenarios, such as black holes in asymptotically flat spacetimes, we would need to look for methods to calculate quantum deviation outside the realm of AdS/CFT.


\acknowledgments
We thank Tron Du, Diego Hofman and Andrea Puhm for helpful discussions. We particularly thank Steve Giddings and the other members of the QuRIOS collaboration for stimulating discussions.
BF is partially supported by Heising-Simons Foundation “Observational Signatures of Quantum Gravity” QuRIOS collaboration grant 2021-2817.


\bibliographystyle{unsrt}
\bibliography{citations}

\end{document}